\documentclass{article}
\usepackage{amsmath, amssymb}
\usepackage{graphicx}
\usepackage[mathscr]{eucal}

\newcommand{\beq}{\begin{equation}}
\newcommand{\eeq}{\end{equation}}
\newcommand{\rf}[1]{(\ref{#1})}
\newcommand{\ba}{\begin{array}}
\newcommand{\ea}{\end{array}}
\newcommand{\dis}{\displaystyle}

\newcommand{\bra}[1]{\langle\,#1\,|}
\newcommand{\ket}[1]{|\,#1\,\rangle}
\newcommand{\p}{\partial}
\newcommand{\ud}{\mathrm{d}}
\newcommand{\pathD}{\!\mathscr{D}}

\newcommand{\x}{{\bf x}}
\newcommand{\y}{{\bf y}}

\newcommand{\q}{{\bf q}}
\newcommand{\pp}{{\bf p}}
\newcommand{\kk}{{\bf k}}
\newcommand{\X}{\mathbf{X}}
\newcommand{\Y}{\mathbf{Y}}

\newcommand{\dirac}[1]{\delta\left(#1\right)}

\newcommand{\ph}{\phi_{p_+}}

\title{Recursive structures in string field theory}
\author{Anton Ilderton \\ Centre for Particle Theory, University of Durham, \\ South Road, Durham DH1 3LE, United Kingdom\\ 
\texttt{anton@cantab.net}}
\date{}
\begin{document}
\maketitle
\abstract{\noindent We derive recursion relations for closed string correlation functions and scattering amplitudes which hold to all orders in perturbation theory.}
\section{Introduction}

Recent years have seen a renewal of interest in recursive calculation of the S-matrix.
This has largely been in the context of gluon scattering using helicity variables \cite{HelicityBook} following \cite{Witten} which is a more efficient and revealing calculational method than the usual LSZ and Feynman diagram expansion. At tree level these recurrence relations are completely understood, along with their connections to twistor theory \cite{treeI}-\cite{treeF} and there is now significant understanding of one loop results \cite{loopsI}-\cite{loopsF}. Recent discovery of an MHV action \cite{Mansfield}-\cite{Morris} opens the way for a systematic exploration of higher loop results.  This program has lead to related results in QED, gravity and other theories \cite{otherI}-\cite{otherF} although results are again mostly limited to one loop.

In a recent paper we presented recurrence relations between QED S-matrix elements \cite{me-QED} (\cite{Nair2} for a similar approach to QCD) which displayed the same basic structure at tree level as the MHV recursion relations but extended simply and explicitly to all loop orders. In this paper we will show that the closed string S-matrix admits a similar recursive expansion using light cone string field theory \cite{SFTI}-\cite{SFTF}. The results will be valid to all orders in the loop expansion.

\section{Recursive expansion of the S-matrix}
The light cone closed string field action is 
\begin{equation}\label{action}
  S[\phi,\phi^\dagger] = \int\pathD X\,\,i\ph^\dagger \p_\tau\ph - \frac{1}{2p_+}\ph^\dagger \hat{h} \ph + S_3[\ph^\dagger, \ph],
\end{equation}
where the string field $\ph$ is a function of light cone time $\tau$, the momentum $p_+$ conjugate to $x_-$ and the 24 transverse co-ordinates $\X$. The 
measure is shorthand for 
\begin{equation*}
  \int\pathD\X^{24}\int\limits_{-\infty}^{\infty}\!\ud\tau\int\limits_0^\infty\!\ud p_+.
\end{equation*}
The operator 
$\hat{h}$ is the first quantised string Hamiltonian,
\begin{equation}
  \hat{h} = \frac{1}{2}\int\limits_0^{2\pi}\!\ud\sigma\,\, 
\X'(\sigma)^2 - \frac{\delta^2}{\delta\X(\sigma)^2}.
\end{equation}
We fix the length of the strings at $2\pi$ rather than the usual $2p_+\pi$. The latter convention naturally lends itself to explicit interaction calculations and though we always consider the interacting theory in this paper we will not need the precise details of the three string vertex. We therefore adopt the former convention which is more familiar from first quantisation.

The generating functional of connected correlation functions, 
$W[J,J^\dagger]$, is given by the functional integral
\begin{equation}
 e^{W[J,J^\dagger]/\hbar} = 
\int\pathD(\phi,\phi^\dagger)\,\,\exp\bigg(\frac{i}{\hbar}S[\phi,\phi^\dagger] + \frac{i}{\hbar}\int\pathD X\ud 
p_+\,\, J^\dagger_{p_+}\phi_{p_+} - J_{p_+}\phi^\dagger_{p_+}\bigg).
\end{equation}
This functional may be expanded in the number of sources $J$, $J^\dagger$ and 
in powers of $\hbar$, corresponding to a loop expansion,
\begin{equation}\begin{split}
  W[J,J^\dagger]=&\sum_{\substack{n,m=1 \\ l=0}} 
\int\prod_{\substack{i=1..n \\ j=1..m}}\pathD(X_i,Y_j)\,\, J^\dagger_{p_{+_1}}\!(X_1)\ldots J^\dagger_{p_{+_n}}\!(X_n)\, J_{q_{+_1}}\!(Y_1)\ldots J_{q_{+_m}}\!(Y_m) \\
  &\times \frac{(i^n)(-i)^m}{n!m!}\bra{0}T \phi_{p_{+}}(X_1)\ldots 
\phi_{p_{+n}}(X_n)\phi_{q_{+1}}^\dagger(Y_1)\ldots \phi_{q_{+m}}^\dagger(Y_m)\ket{0}_L \hbar^L.
\end{split}
\end{equation}
The subscript $L$ indicates the loop order. Subsequent equations will become unwieldy unless we further condense our notation. When unambiguous we will suppress dependencies on the co-ordinates and 
momenta. We will also abbreviate
\begin{equation}
  \bra{0}T \phi_{p_{+}}(X_1)\ldots 
\phi_{p_{+n}}(X_n)\phi_{q_{+1}}^\dagger(Y_1)\ldots \phi_{q_{+m}}^\dagger(Y_m)\ket{0}\equiv \langle 1\ldots 
n;1\ldots m\rangle.
\end{equation}
Then the expansion of the generating functional may be abbreviated to
\begin{equation*}
  W[J,J^\dagger] = \sum\limits_{n,m,L} \int 
\frac{(i^n)(-i)^m}{n!m!}J^{\dagger\,n} J^m \hbar^L\langle 1\ldots n; 1\ldots m\rangle_L.
\end{equation*}
Our goal is to evaluate the expectation value
\begin{equation*}
	\left<\frac{i}{\hbar}S_\text{free}[\phi,\phi^\dagger]\right>,
\end{equation*}
which we may write in terms of functional derivatives with respect to 
the sources,
\begin{equation}\label{expect-op}
	i\hbar\int\pathD\X\int\!\ud(p_+,\tau)\,\, \frac{\delta}{\delta 
J_{p_+}(\X,\tau)}\bigg(i\p_\tau - \frac{\hat{h}}{2p_+}\bigg)\frac{\delta}{\delta 
J^\dagger_{p_+}(\X,\tau)} e^{W[J,J^\dagger]/\hbar}.
\end{equation}
We are not interested in free correlators so we separate $W$ into
\begin{equation}
	W = \int J^\dagger\bigg(\p_\tau + i\frac{\hat{h}}{2p_+}\bigg)^{-1}J + 
\widetilde{W}[J,J^\dagger],
\end{equation}
where $\widetilde{W}$ contains correlation functions of the interacting 
field. This decomposition reveals a divergence in the expectation value coming from the free part. This is because (\ref{expect-op}) involves operators evaluated at coincident spacetime 
points and therefore stands in need of regularisation. To do this we 
will insert the operator $\mathcal{T_\epsilon}$ into the free action 
acting on $\phi^\dagger$, the effect of which is to shift the arguments of 
the field according to
\begin{equation*}
 \mathcal{T}_\epsilon\phi^\dagger_{p_+}[\X,\tau] = \phi_{p_+ + 
\epsilon}^\dagger [\X+\boldsymbol{\epsilon},\tau+\epsilon].
\end{equation*}
where $(\epsilon,\boldsymbol{\epsilon}, \epsilon_+)$ is some small 
change in position and momentum. This removes the divergent term. 
Evaluating (\ref{expect-op}) and letting $\epsilon\rightarrow 0$ we find
\begin{equation}\begin{split}\label{equate1}
  e^{-W/\hbar}\left<\frac{i}{\hbar}S_\text{free}[\phi,\phi^\dagger]\right> = &-\frac{1}{\hbar}\int\frac{\delta \widetilde{W}}{\delta J}\bigg(\p_\tau +i \frac{\hat{h}}{2p_+}\bigg)\frac{\delta \widetilde{W}}{\delta J^\dagger} - \int \frac{\delta}{\delta J}\bigg(\p_\tau +i \frac{\hat{h}}{2p_+}\bigg)\frac{\delta \widetilde{W}}{\delta J^\dagger} \\
  &-\int J\bigg(\p_\tau + i\frac{\hat{h}}{2p_+}\bigg)^{-1}J^\dagger - \frac{1}{\hbar}\int J\frac{\delta \widetilde{W}}{\delta J} - \frac{1}{\hbar}\int J^\dagger\frac{\delta \widetilde{W}}{\delta J^\dagger}.
\end{split}\end{equation}
The effect of the final two terms is simply to count the number of 
string fields so that the $\phi^n$, $\phi^{\dagger m}$ 
correlator in the perturbative expansion of 
$\widetilde{W}$ is multiplied by $-(n+m)$. Including the regularisation we may also write the expectation value as
\begin{equation}\label{with zeta}
  \left<\frac{i}{\hbar}S_\text{free}[\phi,\phi^\dagger]\right>=\frac{\p}{\p\zeta}\bigg|_{\zeta=0} 
\int\pathD(\phi,\phi^\dagger)\,\,e^{i\widetilde{S}/\hbar + i\int\! J^\dagger\phi - J\phi^\dagger/\hbar},
\end{equation}
where the action $\widetilde{S}$ is defined by
\begin{equation*}
\widetilde{S} = \int\!\ph^\dagger \bigg(i\p_\tau - \frac{\hat{h}}{2p_+} 
\bigg)(1+\zeta\mathcal{T}_\epsilon)\ph + S_3[\ph^\dagger, \ph].
\end{equation*}
The functional integral in $\rf{with zeta}$ is given by the usual 
Feynman diagram expansion of the generating functional except the propagator 
$G$ is replaced by
\begin{equation*}
  (1+\zeta\mathcal{T}_\epsilon)^{-1}G.
\end{equation*}
The inverse operator may be defined as a formal power series in 
$\zeta$. The derivative, as $\zeta$ goes to zero, multiplies the exponential 
of $W/\hbar$ by a sum of Feynman diagrams derived from the usual sum 
over connected diagrams as follows. Each connected diagram is separated 
into $E+I$ similar diagrams, where $E$ ($I$) is the number of external 
(internal) lines. In each of these diagrams the $\zeta$ derivative acts on one propagator. As $\zeta$ goes to zero this propagator, $G$, is replaced by $-\mathcal{T}_\epsilon G$.

As we let the regulator $\epsilon\rightarrow 0$ we therefore recover 
the sum over connected diagrams where each is multiplied by a factor of 
$-E-I$. Since the interaction is cubic the number of internal lines in a 
connected diagram is a function of the number of external lines and the loop 
order $L$ given by $I^E_L:=E+3(L-1)$. The expectation value may then 
be written
\begin{equation}\begin{split}
e^{-W/\hbar}&\left<\frac{i}{\hbar}S_\text{free}[\phi,\phi^\dagger]\right> = -\int J\bigg(\p_\tau + i\frac{\hat{h}}{2p_+}\bigg)^{-1}J^\dagger 
\\
  & -\sum\limits_{n,m,L} \int \frac{(i^n)(-i)^m}{n!m!}J^{\dagger\,n} 
J^m \hbar^l\langle 1\ldots n;1\ldots m\rangle_L(n+m+I^{n+m}_L) \label{equate2}.
\end{split}\end{equation}
Equating the expressions \rf{equate1} and \rf{equate2} we find
\begin{equation}\begin{split}
  \sum\limits_{n,m,L} \int &\frac{(i^n)(-i)^m}{n!m!}J^{\dagger\,n} J^m 
\hbar^l\langle 1\ldots n;1\ldots m\rangle_L I^{n+m}_L = \\
  & \int\frac{\delta \widetilde{W}}{\delta J}\bigg(\p_\tau +i 
\frac{\hat{h}}{2p_+}\bigg)\frac{\delta \widetilde{W}}{\delta J^\dagger} + \hbar\int 
\frac{\delta}{\delta J}\bigg(\p_\tau +i \frac{\hat{h}}{2p_+}\bigg)\frac{\delta 
\widetilde{W}}{\delta J^\dagger}.
\end{split}\end{equation}
The free part of the generating functional has dropped out of our 
equations. Equating by order in $J$, $J^\dagger$ and $\hbar$ we find
\begin{equation}\begin{split}
  &I^{n+m}_L\langle 1\ldots n ; 1\ldots m \rangle_L = \int\pathD X\,\, 
\bigg(\p_\tau +i\frac{\hat{h}(\X)}{2p_+}\bigg)\langle 1\ldots n, X ; 
1\ldots m, Y\rangle_{L-1}\bigg|_{Y=X} \\
	& + \int\pathD X\,\,\hspace{-10pt}\sum_{\substack{\sigma \\ r=0..n \\ 
s=0..m \\ L'=0..L}} R^{n,m}_{r,s} \langle 1\ldots n-r ; 1\ldots s, 
X\rangle_{L'} \bigg(\p_\tau +i\frac{\hat{h}(\X)}{2p_+}\bigg)\langle 1\ldots 
r, X ; 1\ldots m-s\rangle_{L-L'},
\end{split}\label{RESULT}\end{equation}
where $R^{n,m}_{r,s} = 1/(r!s!(n-r!)(m-s!)$. The sum over $\sigma$ is 
over all distributions of the external data $\{X_r,Y_r\}$ between the 
two correlation functions. We claim that these relations are recursive in the number of string field operators and in the loop order, but the initial and final terms of the sums may appear to spoil this. 
On the left hand side is a correlator of $n+m$ 
fields at loop level $L$, while the right hand side is built from pairs of correlators nearly all of which are for $<n$ fields at $<L$ loops. However when $r=0$ and $s=m-1$ one correlator contains a total of $m+n$ fields and may 
be of loop level $L$. Observe though that this is coupled to a correlator of two 
fields at loop level 0, which is just the free propagator. By construction this 
object does not appear in $\widetilde{W}$ and this term vanishes from the sum. Therefore, although the right hand side contains 
correlators of $n+m$ fields they are of loop order at most $L-1$. When $r=n$ and $s=0$ the right hand side 
contains one field correlators (tadpoles and higher loop contributions) 
coupled to $n+m+1$ field correlators. When the latter is of loop order 
$L$ the one point function is tree level. Since there are no tree level 
tadpoles, by definition, this term vanishes and correlators of $n+m+1$ 
fields are of loop order at most $L-1$. Equation (\ref{RESULT}) therefore gives a genuinely recursive expansion of the partition function in the number of correlated fields and the loop order.

It is now a short step to go to a similar expansion of the S-matrix. One method is 
to repeat the above analysis with the sources $J$ and $J^\dagger$ 
replaced by $J^\dagger(\p_\tau + i\hat{h}/2p_+)$ and $J(-\p_\tau + 
i\hat{h}/2p_+)$. Differentiation with respect to the Fourier modes of $J$ and 
$J^\dagger$ brings down the Fourier transform of correlation 
functions with amputated external legs, i.e. (generally off-shell) S-matrix 
elements. The calculation proceeds as before. Alternatively, we may 
directly amputate and Fourier transform equation (\ref{RESULT}).

The creation and annihilation modes may be extracted from the string 
field in the usual way (see appendix),
\begin{equation*}\begin{split}
  A^{\dagger\text{ in}}_{p_+,\pp,\{l\}} &= \int\pathD \X\,\, f_{\{l\}}(\X) e^{-iE_-\tau + i\pp.\x}\phi^\dagger_{p_+}(X), \\
A^{\text{out}}_{p_+,\pp,\{l\}} &= \int\pathD \X\,\, f_{\{l\}}(\X) e^{iE_-\tau - i\pp.\x}\phi_{p_+}(X).
\end{split}
\end{equation*}
The LSZ reduction formula for S-matrix elements follows using the usual 
methods. Denoting external data $\{p_{+r},E_-, \pp_r\}$ by simply 
$r$, the S-matrix element
\begin{equation*}
 \delta(E_{-\text{in}} - E_{-\text{out}})\delta(\pp_\text{in} - 
\pp_\text{out})\delta(p_{+\text{in}} - p_{+\text{out}})\mathcal{A}(1\ldots n | 
1\ldots m)_L 
\end{equation*}
is given by a product of terms
\begin{equation}\label{act1}
  \int\!\ud\tau_r\!\int\pathD \X_r\,\, e^{i(E_-(l_r,\pp_r,p_{+r})\tau_r - \pp_r.\x_r)} 
f_{\{l\}_r}(X_r) \bigg(\p_\tau +i\frac{\hat{h}(\X_r)}{2{p_{+r}}}\bigg)
\end{equation}
for each `out' state and
\begin{equation}\label{act2}
 \int\!\ud\tau'_r\!\int\pathD \Y_r\,\, e^{-i(E_-(l_r,\q_r,q_{+r})\tau'_r - \q_r.\x_r)} 
f_{\{l\}_r}(X_r) \bigg(-\p_{\tau'} +i\frac{\hat{h}(\Y_r)}{2{q_{+r}}}\bigg)
\end{equation}
for each `in' state, all acting on the correlator
\begin{equation*}
\bra{0}T \phi_{p_{+}}(X_1)\ldots 
\phi_{p_{+n}}(X_n)\phi_{q_{+1}}^\dagger(Y_1)\ldots \phi_{q_{+m}}^\dagger(Y_m)\ket{0}_L.
\end{equation*}
We apply these operators to \rf{RESULT}. For the integrated (internal line)
data appearing in the correlation functions we may use the spectral 
decomposition of the propagator in order to introduce the
prefactors \rf{act1} and \rf{act2} associated to the off-shell string fields;
\begin{equation}
 G_{p_+,q_+}(\X,\tau; \Y, \tau') = 
i\dirac{p_+-q_+}\sum\limits_{\{l\}}\int\!\frac{\ud^{24}\q\ud q_-}{(2\pi)^{25}}\,\,e^{-iq_-(\tau-\tau') + 
i\q(\x-\y)}\frac{f_{\{l\}}(\X)f_{\{l\}}(\Y)}{q_- - \frac{h(\q, 
\{l\})}{2p_+}}.
\end{equation}
We then find that the scattering of $m$ to $n$ strings is given by
\begin{equation}\label{RESULT2}
\begin{split}
  &I^{n+m}_L\mathcal{A}(1\ldots n | 1\ldots m)_L = 
\sum\limits_{\{l\}}\int\frac{i\ud^\mu k}{k_- - \frac{h(\kk,\{l\})} {2k_+}}\mathcal{A}(1\ldots n , \{k,\{l\}\}|1\ldots m, \{k,\{l\}\})_{L-1}\\
&+\sum R^{n,m}_{r,s}\mathcal{A}(1\ldots n-r|1\ldots s, \{q, \{l\}\})_{L'}\,\frac{i}{q_- -\frac{h(\q,\{l\})} {2q_+}} \\
&\hspace{2in}\times \mathcal{A}(1\ldots r, \{q,\{l\}\}|1\ldots m-s)_{L-L'}.
\end{split}
\end{equation}
In the first line the measure is $\ud^\mu k = 
(2\pi)^{-25}\ud^{24}\kk\,\ud k_+\,\ud k_-$. In the second line the sum is as in \rf{RESULT} but 
now includes a sum over all possible excitations $\{l\}$ of the string. 
The momentum $q = \{q_+, \q, q_-\}$ is forced, by the overall delta 
functions in the scattering amplitudes, to conserve momentum between pairs 
of amplitudes. If we restrict to tree level then the integrated terms in (\ref{RESULT}) and (\ref{RESULT2}) go out. Beyond tree level these terms remain and describe how $L$-loop amplitudes of $E$ external
strings receive contributions from $L-1$ loop amplitudes of $E+2$ 
external strings - the two additional external legs are sewn together using a 
propagator to form a closed loop.

\section{Discussion}
We have derived recurrence relations between scattering amplitudes of closed strings using light cone string field theory. We emphasise that in contrast to other approaches our recursion relations hold explicitly to all loop orders.

In one sense this recursive structure is more natural in string theories than in 
particle theories where worldsheets may be joined together to form Feynman diagrams using Carlip's sewing method \cite{Carlip}. Two worldsheets are sewn together by acting on two boundaries with the first quantised Hamiltonian (which gives the correct moduli space measure on the sewn worldsheet), inserting a closed string propagator between the boundaries and integrating over the common boundary data. Our results are effectively an application of this method, albeit in a Lagrangian formalism; pairs of amplitudes 
are sewn together using the inverse of the two point function, which is the inverse of the first quantised Hamiltonian.

We have not needed the explicit form of the three string interaction. Any interaction of the form
\begin{equation*}
  g\int\! V\phi^2\phi^\dagger + V^\dagger\phi^{\dagger 2}\phi,
\end{equation*}
with $V$ some integral kernel (such as the usual delta functional in all variables) will lead to the same recursion relations -- we may in fact any number of cubic terms with any number of derivatives acting on the fields, provided we satisfy physical constraints such as reparametrisation invariance. The chosen interaction kernel provides the initial data for constructing amplitudes recursively. We note that our method extends to non-flat spacetimes, for example the Plane Wave background \cite{PP}. The recurrence relations hold as in \rf{RESULT} and \rf{RESULT2}, with $\hat{h}$ replaced by the plane wave Hamiltonian,
\begin{equation*}
  \hat{h}_{PW} = \hat{h}_\text{flat}+\int\limits_0^{2\pi}\!\ud\sigma\,\, \mu^2 p_+^2\X(\sigma)^2.
\end{equation*}
Formally, our arguments appear to extend immediately to any supersymmetric string field theory with a cubic vertex, e.g. the Heterotic and Type IIAB strings -- the differences would amount to replacing the quadratic form in \rf{RESULT} and \rf{RESULT2} with that appearing in the superstring action. This would remove the tachyon divergences of the bosonic string which we have not treated here. Superstring field theory suffers however from contact divergences \cite{Contact} \cite{Contact2} \cite{recentContact} which are absent in 
the bosonic case. The regularisation of these divergences generally requires the addition of counter terms which are of higher than cubic order. These higher order terms would considerably complicate the form of the recursion relations we have derived. An interesting question for future study is whether is it possible to regulate contact term divergences while maintaining the simple structure of the bosonic relations.
\appendix
\section{Conventions}
For completeness we present our conventions. The equations of motion 
which follow from the action \rf{action} are
\begin{equation*}
  \big(i\p_\tau - \frac{1}{2p_+}\hat{h}\big)\ph =0,\quad \big(-i\p_\tau - \frac{1}{2p_+}\hat{h}\big)\ph^\dagger =0.
\end{equation*}
To solve these equations we first expand the transverse co-ordinates in 
a suitable basis,
\begin{equation*}
   {\dis \X(\sigma) = \frac{\x}{\sqrt{2\pi}}+ 
\frac{1}{\sqrt{\pi}}\sum\limits_{n=1}^\infty \x_n\cos n\sigma  + \bar{\x}_n\sin n\sigma },
\end{equation*}
\begin{equation*}
  \implies \frac{\delta}{\delta\X(\sigma)} = 
\frac{1}{\sqrt{2\pi}}\frac{\p}{\p\x} + \frac{1}{\sqrt{\pi}} \sum\limits_{n=1}^\infty 
\frac{\p}{\p\x_n}\cos n\sigma + \frac{\p}{\p\bar{\x}_n}\sin n\sigma.
\end{equation*}
This allows us to expand $\hat{h}$ in modes such that the equations of motion 
become partial differential equations which we solve by separation of variables. With a slight abuse of 
notation, the separated equations are
\begin{eqnarray}
  \nonumber i\p_\tau\ph &=& E_-\ph, \\ \nonumber \p_0^2\ph &=& -\pp^2\ph, \\
  \frac{1}{2}\bigg(\frac{\p^2}{\p (x_n^i)^2} -n^2 (x^i_n)^2\bigg)\ph 
&=& -E_{n,i}\ph, \quad\text{for $i=1\ldots24$}, \label{hermite1}\\
  \nonumber\frac{1}{2}\bigg(\frac{\p^2}{\p (\bar{x}_n^i)^2} -n^2 
(\bar{x}^i_n)^2\bigg)\ph &=& -\bar{E}_{n,i}\ph, \quad\text{for 
$i=1\ldots24$},\label{hermite2} 
\end{eqnarray}
with the constraint
\begin{equation}
  E_- =\frac{1}{2p_+}\bigg(\frac{1}{2}\pp^2 - \sum\limits_{n,i} E_{n,i} 
+ \bar{E}_{n,i}\bigg).
\label{constraint}\end{equation}
We recognise (\ref{hermite1}) as Hermite's equation, for well behaved solutions to which we must have $2E_n/n = 2l+1$ for $l$ a non-negative integer which we call the level. The 
solution of level $l$ is
\begin{equation}
  H_{l_{n,i}}(\sqrt{n}x^i_n)e^{-n(x^i_n)^2/2},
\end{equation}
where $H_l$ is a Hermite polynomial. This means that for each mode of the string, describing a quantum 
harmonic oscillator of frequency $n$, in each dimension transverse $i$, 
there are an infinite number of solutions to the equations of motion 
representing the infinite number of excitations of the oscillator, labelled 
by $l_{n,i}$. The constraint equation \rf{constraint} becomes
\begin{equation}
  2p_+E_- -\pp^2 - \sum\limits_{i=1}^{24}\sum\limits_{n=1}^\infty n 
l_{n,i} + n\bar{l}_{n,i} + n =0.
\end{equation}
The sum over the final term diverges, and must be regulated. Zeta 
function regularisation implies 
\begin{equation}
  E_-=\frac{1}{2p_+}\bigg(-2 + \sum\limits_{n,i} 
nl_{n,i}+n\bar{l}_{n,i}\bigg) = \frac{L_0+\bar{L}_0-2}{2p_+},
\end{equation}
which is proportional to the first quantised string Hamiltonian 
$L_0+\bar{L}_0$ which treats $l$ as the number of excited string oscillators of 
frequency $n$. The full solution to the equations of motion is then
\begin{equation}
  \ph(\tau,\X) = \int\!\frac{\ud^{24} \pp}{(2\pi)^{24}} 
\sum\limits_{\{l\}} A_{\pp,p_+,\{l\}} e^{-iE_-\tau}e^{i\pp.\x_0}f_{\{l\}}(\X),
\label{thestring2}\end{equation}
where the functions $f_{\{l\}}$ are defined by
\begin{equation}
 f_{\{l\}}(\X):=\prod\limits_{n,i} 
\frac{H_{l_{n,i}}(\sqrt{n}x_n^i)H_{\bar{l}_{n,i}}(\sqrt{n}\bar{x}_n^i)}{(2^{l+\bar{l}} l!\,\bar{l}!)^{1/2} 
(\pi/n)^{1/2}}e^{-n(x_n^i)^2/2}e^{-n(\bar{x}_n^i)^2/2}.
\end{equation}
The product over $n$ and $i$ is the product over the separated Hermite 
solutions for each string mode $x_n$ in dimension $i$. There is then a 
sum over all possible combinations of assigning level values to each of 
these solutions, representing the infinite number of independent 
solutions to the equations of motion. We have included a normalisation factor 
in the Hermite polynomials, which obey
\begin{equation}
  \int\!\ud x\,\, e^{-nx^2}H_{l_1}(\sqrt{n}x)H_{l_2}(\sqrt{n}x) = 2^l 
l!(\pi/n)^{1/2}\delta_{l_1, l_2}.
\end{equation}

\end{document}